\documentstyle{l-aa}
\begin{document}
\input{psfig} \def\capo{\\ \indent} \def\cali{\par\noindent}
\def\bequ{\begin{center}\begin{equation}}
\def\fequ{\end{equation}\end{center}} \def\com#1{{\large{\bf #1}}}
\title{Cosmology with Sunyaev-Zeldovich observations from space}

\author{N. Aghanim\inst{1} \and A. De Luca\inst{1} \and F. R. Bouchet\inst{2}
\and R. Gispert\inst{1} \and J. L. Puget\inst{1}}

\offprints{ }

\institute{IAS-CNRS, Universit\'e Paris XI, Batiment 121, F-91405 Orsay Cedex
\and IAP-CNRS, 98 bis, Boulevard Arago, F-75014 Paris}

\date{Received date; accepted date} \maketitle

\begin{abstract}

In order to assess the potential of future microwave anisotropy space
experiments for detecting clusters by their Sunyaev-Zeldovich (SZ) thermal 
effect, we have
simulated maps of the large scale distribution of their Compton parameter
$y$ and of the temperature anisotropy $\Delta T/T$ induced by their proper
motion. Our model is based on a predicted distribution of clusters per unit
redshift and flux density using a Press-Schecter approach (De Luca et
al. 1995).

These maps were used to create simulated microwave sky by adding them to 
the
microwave contributions of the emissions of our Galaxy (free-free, dust and
synchrotron) and the primary Cosmic Microwave Background (CMB) anisotropies
(corresponding to a COBE-normalized standard Cold Dark Model scenario). In
order to simulate measurements representative of what current technology
should achieve, ``observations'' were performed according to the instrumental
characteristics (number of spectral bands, angular resolutions and detector
sensitivity) of the COBRAS/SAMBA space mission. These observations were
separated into physical components by an extension of the Wiener filtering
theory (Bouchet et al. 1996). We then analyzed the resulting $y$ and $\Delta
T/T$ maps which now include both the primary anisotropies and those
superimposed due to cluster motions.
A cluster list was obtained from the recovered $y$ maps, and their profiles
compared with the input ones. Even for low $y$-values, the input and output
profiles show good agreement, most notably in the outer parts of the profile
where values as low as $y \simeq 3.\,10^{-7}$ are properly mapped. We also
construct and optimize a spatial filter which is used to derive the accuracy
on the measurement of the radial peculiar velocity of a detected cluster. We
derive the accuracy of the mapping of the very large scale cosmic
velocity field obtained from such measurements.

\end{abstract}
\section{Introduction}

Future space experiments should allow to built a large
statistically homogeneous sample of clusters of galaxies detected by the 
spectral
signature of CMB photons scaterring off the free electrons of the hot
intra-cluster gas. This would be of great cosmological interest, allowing to 
constrain
the cosmological scenarios of large scale structure formation and evolution,
as well as the gas history (Colafrancesco \& Vittorio1994, Barbosa et al. 
1996).

The measurement of the peculiar velocity of clusters of galaxies could be an
important tool to study the large scale velocity field of the universe. This
in turn provides a unique opportunity for probing the underlying mass
distribution. Thus one can probe fairly directly the primordial spectrum and
further constrain the various cosmological models.  The direct determination
of the peculiar velocity (by independent redshift and distance determination)
is observationally difficult and time-consuming. Still, several groups have
succeeded in measuring the bulk, volume-averaged, peculiar velocity field, in
our neighborhood, on scales $r = 10\, h^{-1}$ to $60\, h^{-1}$ Mpc (Aaronson
et al. 1986, Collins et al. 1986, Dressler et al. 1987, Strauss
\& Willick 1995 (for a recent review)).

Another method is nevertheless possible. It relies on combining the
informations from the Sunyaev-Zeldovich (SZ) effects (thermal and kinetic)
which create secondary temperature fluctuations (Zeldovich \& Sunyaev 1969,
Sunyaev \& Zeldovich 1972, 1980). The combination of the measurements of the
two effects give a fairly direct determination of the peculiar radial velocity
of clusters of galaxies (Sect.  2.2, Eq. 3), although a good estimate of the
peculiar velocity requires a precise measurement of both SZ effects and of the 
gas temperature. A crucial
advantage of such a combination is that the SZ effects do not decrease in
brightness with distance. Therefore measurements of both effects for a large
number of clusters could give precise estimates of the large scale velocity
field distribution.  \par The temperature  fluctuations generated by the 
kinetic SZ 
effect have the same spectral
signature than the primordial anisotropies of the CMB. For cluster
velocity determinations, the primordial temperature fluctuations of the CMB
act as a contaminating source of the SZ kinetic effect. In addition, one
must also take into account all other sources of contamination that might
spoil the measurement, from resolved sources (other clusters, galaxies ...),
unresolved ones (galactic synchrotron, free-free emission ...) or instrumental
noise.  \par

Some measurements of the SZ thermal effect, on known clusters, have already 
been
made in the wavelength range where the SZ effect creates a decrement in the
CMB intensity ($\lambda>1.3$ mm) (see Rephaeli 1995 and references therein).
\par In the context of the feasibility study of future space projects such as
the COBRAS/SAMBA mission\footnote{For a complete description of the
COBRAS/SAMBA project, see ESA document D/SCI(96)3.}, of the European Space
Agency, dedicated to the CMB observations together with other cosmological
targets (SZ detection in clusters, primordial galaxies, ...), a complete
simulation of the astrophysical processes, the expected instrumental
characteristics of the satellite and of the separation of these processes was
performed (Bouchet et al. 1996). This simulation was used as a tool to
constrain and quantify the capabilities of the satellite. In this paper, we
focus on the capabilities of such a mission for the detection of the SZ
thermal effect of clusters of galaxies, the imaging of the clusters and
finally the measurement of their peculiar velocities.  \par

Section 2 introduces briefly the formalism of the thermal and kinetic SZ
effects, while we present in section 3 the method we used to generate SZ
effect maps. Section 4 assesses the capabilities of a space experiment like
COBRAS/SAMBA to detect clusters, and we evaluate, in section 5, the 
accuracy of
the peculiar velocity determination of individual clusters. We finally derive
the expected accuracy on the rms dispersion of the large scale velocity field
in section 6.  Section 7 summarizes the main results and conclusions of the
paper.
\section{Sunyaev-Zeldovich thermal and kinetic effects}
\subsection{Thermal Sunyaev-Zeldovich effect}
The thermal SZ effect is the inverse Compton scattering of CMB photons by free
electrons in the hot intra-cluster medium. Since the number of photons is
conserved, their spectrum is just shifted on average to higher frequencies.  
\\ This effect is
characterized by the comptonization parameter $y$, which depends only on the
cluster's electronic temperature and density ($T_e$, $n_e$):
\[
y=\frac{k\sigma_T}{m_ec^2}\int T_e(l)n_e(l) dl
\]
\noindent
where $k$ is the Boltzmann constant, $\sigma_T$ the Thomson cross section,
$m_ec^2$ the electron rest mass energy and $l$ is the distance along the line 
of
sight.  When the intra-cluster gas is isothermal ($T_e(l)=T_e=const$), $y$ is
expressed as a function of the optical depth $\tau$ ($\tau=\sigma_T\int n_e(l)
\,dl$): \bequ y=\tau~\frac{kT_e}{m_ec^2}.
\label{eq:yparam}
\fequ The relative monochromatic intensity variation of the CMB due to the SZ
thermal effect is given by
\[
\frac{\Delta I_{\nu}}{I_{\nu}}=y\cdot f(x)
\]
\noindent
where $x$ is the adimentionnal frequency $x=h\nu/kT_{CMB}$ ($h$ denotes the
Planck constant, $T_{CMB}$ the CMB temperature, and $\nu$ the frequency),
$I_{\nu}$ is the intensity of the CMB (black body emission) and $f(x)$ is the
spectral form factor given by:
\[
f(x)=\frac{xe^x}{(e^x-1)}\left[x\left(\frac{e^x+1}{e^x-1}\right)-4\right].
\]
\subsection{Kinetic Sunyaev-Zeldovich effect} 
If a cluster has a radial peculiar velocity $v_r$, another relative intensity
variation of the CMB due to the Doppler first-order effect is added.  It is
given by: \bequ \frac{\Delta I_{\nu}}{I_{\nu}}=- \frac{v_r}{c}\tau\cdot a(x)=
\left(\frac{\Delta T}{T}\right)_{SZ}\cdot a(x), \fequ where $a(x)$ is the
spectral form factor for the kinetic effect, given by:
\[
a(x)=x\frac{e^x}{e^x-1},
\]
\noindent
and $\tau$ is the optical depth. The effect is positive for clusters moving
towards the observer (i.e. with negative velocities).  \par
\noindent
The intensity fluctuation induced by the SZ kinetic effect has the same
spectral shape as the primordial ones (equivalent to a temperature
fluctuation).  \par\noindent The temperature variation due to the kinetic SZ
effect can be written as a function of the $y$ parameter
(defined in Eq. \ref{eq:yparam}). The combination of the two SZ effects gives 
the
expression of the radial velocity of the cluster: \bequ
v_r=-c\,\frac{kT_e}{m_ec^2}\frac{(\Delta T/T)_{SZ}}{y}.
\label{eq:vr}
\fequ
At high frequencies, the expressions given above are not
accurate enough and relativistic calculations of the thermal effect are needed 
in most cases (Rephaeli
1995 and references therein). A relativistic treatment introduces differences
in the relative intensity variation and a shift of the crossover frequency. 
The corrections depend on both the temperature of the intracluster medium and  
the frequency. In this paper, we have restricted our study to the 
nonrelativistic treatment as a ``text book'' case in order to adress the 
capabilities of a space mission in measuring the SZ effect on clusters of
galaxies. The corrections for the relativistic case will 
have to be taken into account when dealing with real data. 
\section{Simulations}
\subsection{ Cluster model} 
We model the $y$ profile of a single resolved cluster according to a King
model (King 1966).  The main physical characteristics of the cluster are
$n_e(R)$ and $T_e(R)$, respectively the electronic density and temperature
distributions given as functions of the distance $R$ to the center of the
cluster.  Hereafter, we use a hydrostatic isothermal model with a spherical
geometry (Cavaliere \& Fusco Femiano 1978, Birkinshaw, Hugues \& Arnaud
1991). For the density distribution, this gives:
\[
n_e(R)=n_{e0}\left[1+\left(\frac{R}{R_c}\right)^2\right]^{-\frac {3\beta
}{2}},
\]
\noindent
where $n_{e0}$ is the central electronic density, $R_c$ is the core radius and
$\beta$ is a parameter of the model which represents the ratio of the kinetic
energy per unit mass in the galaxies to the one in the gas. We take
$\beta=2/3$, as indicated by both numerical simulations (Evrard 1990) and
X-ray surface brightness profiles (Jones \& Forman 1984, Edge \& Stewart
1991).  The hypothesis of isothermality is rather well confirmed by the X-rays
observations of the ASCA satellite (Mushotzky 1994).

In this specific model for the gas distribution, the integrated $y$ profile
of the cluster is given by:
\[
y \propto \left[ 1+ \left( \frac{\theta}{\theta_c}\right)^2 \right]^{-1/2},
\]
\noindent
$\theta$ and $\theta_c$ being respectively the angular distance to the center 
of the cluster and core radius.
The Full Width at Half Maximum (FWHM) of the profile and the core radius of 
the cluster are
related by $FWHM=3.5 \times \theta_c$.

\par

For the time evolution of the temperature $T_{e}$ and core radius $R_c$, we use
the model of Bartlett and Silk (1994) in which the main parameters of the 
cluster evolve as suggested by the self-similarity arguments (Kaiser 1986), 
with a parameter standing for the negative evolution of the number of clusters
indicated by the X-ray observations. The normalization of the temperature and 
core radius is made using the A665 ASCA 
data (Yamashita 1994).
\subsection{The maps}
We simulate $500\times500$ pixels ($12.5^{\circ}\times 12.5^{\circ}$) maps
of the clusters, in terms of their $y$ parameters for the thermal SZ effect
and $\Delta T/T$ maps of the same size for the kinetic SZ effect.  The
distribution of clusters per unit of redshift, solid angle and flux density
interval was computed (De Luca et al. 1995) using a Press-Schechter mass
function (Press \& Schechter 1974) normalized to X-ray and optical data
(Bahcall \& Cen 1993). The total number of sources in each map is drawn using
a Poisson distribution, the position of each cluster being also assigned at
random, i.e. the effect of spatial correlations are not taken into account.  
\par\noindent
The counts in De Luca et al. (1995) are obtained using a formalism similar
to the one adopted in 
Bartlett \& Silk (1994) taking into account the negative evolution in time 
suggested by X-ray observations. 
\par
A discrimination between resolved and unresolved clusters is made according to
their spatial extent. We give the gas distribution
within clusters a finite extent with a maximum radius $R_{max}$. In fact, the
hydrodynamic isothermal model is in good agreement with X-ray observations
over 5 core radii (Markevitch et al. 1992), other observations (Henriksen
\& Mushotzky 1985) show that in some cases the gas in clusters is still seen
up to 10 core radii. This seems to be also the case in the ASCA data
(Mushotzky 1994). These sizes are the ones estimated from the X-ray 
observations.
Since the X-ray emisssion is proportional to $n_e^2$, the observations are 
strongly biaised 
towards the center of the clusters and tend to minimize the extent of 
the
clusters. Recent simulations of cluster formation (Evrard et al. 1996) show 
that
the virialization limit for the clusters takes place at 2 to 3 Mpc from the 
center. 
For typical values of the core radius (0.1 to 0.2 Mpc) the dynamical extent
of the clusters is thus of the order of 10 to a few tens of core radii. 
Because the SZ effect is
proportional to the $n_e$, it is therefore very sensitive to the outer parts 
of the density profile where most of the mass rests. In order to investigate 
the case where the clusters extend 
beyond the estimations from X-rays, 
we assume, for the simulated maps, that the maximum extent of a
cluster is $20\theta_c$ with a profile going down to zero for
$\theta>10\theta_c$. In the maps, the unresolved sources are thus the ones
for which $20\theta_c < 1.5'$, where $1.5'$ is the simulation pixel size. In
that case, we assign to the corresponding pixel an integrated $y$ parameter,
noted Y (see Eq. 4).  When the sources are spatially resolved ($20\theta_c >
1.5'$), we compute their $y$ profile using the above $\beta$-model for the gas
(Sect 3.1 and appendix A).  \par

To generate the $\Delta T/T$ maps arising from the SZ kinetic effect, the
radial peculiar velocity $v_r$ of each cluster is drawn randomly from an
assumed Gaussian velocity distribution with standard deviation today
$\sigma_0=400\, km/s$ (Faber et al. 1993). The time evolution of the standard
deviation is followed according to linear perturbation theory. We also assume
no correlations in the velocity distribution. Using the same procedure than
above concerning the source extents, we compute $\Delta T/T$ profiles for the
resolved clusters (Appendix A).  \\ We found that this simple model based on
De Luca et al. (1995) modeling  for the SZ source counts turns out to be 
in rather good agreement with
Bond \& Myers' (1996) more sophisticated simulations, based on the ``peak
patch'' algorithm, at least for statistical quantities such as the $rms$
values of both the $y$ parameter and $\Delta T/T$. This justifies the
approximations made concerning correlations.
\section{Detection and mapping of distant clusters}
It is now possible, due to recent technological improvements, to design a new
generation of satellites dedicated to CMB observations at small angular scales
(few arcminutes to few degrees) with very high sensitivities, allowing the
detection of temperature fluctuations at a level of $\Delta T/T \simeq
10^{-6}$. Measuring primordial temperature fluctuations at this level of
accuracy requires the ability to detect secondary fluctuations such as the
ones associated with the SZ effect. Therefore one of the byproducts of such a
survey of the CMB will certainly be a new catalogue of clusters of galaxies.
\\ The physics of cluster formation and the history of gas virialization are
still almost unknown. The SZ effect measurement provides a new method of
observing clusters which is potentially more powerful than X-ray observations
for the search of clusters at high redshifts.
\begin{figure}
\psfig{file=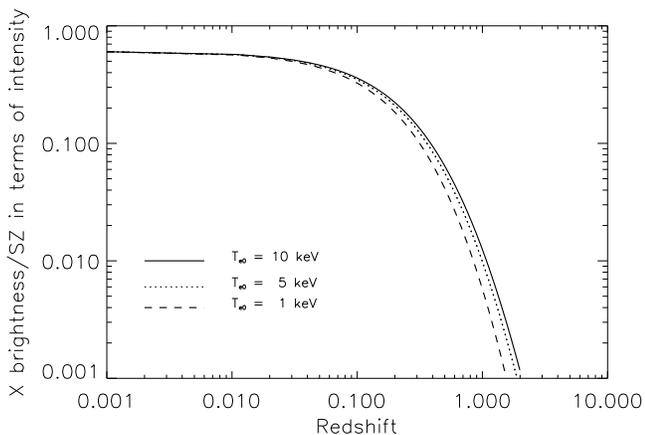,width=\columnwidth}
\caption[ ]{Ratio of the X brightness to the intensity of the SZ thermal
effect as a function of the redshift, for three temperatures of the cluster.}
\end{figure}
\cali
Figure 1 gives the ratio of the X-ray brightness of a galaxy cluster to the 
intensity of its SZ
thermal effect, as a function of the redshift. It shows
that at $z=1$ the X-ray brightness drops by a factor greater than 30 compared
to the SZ intensity, indicating that SZ observations should be very powerful
to detect high $z$ clusters.  \par

In the context of CMB observations at small angular scales with high
sensitivities, the problem of the detectability of clusters among the other
astrophysical components is essential to evaluate how many clusters could be
found in a sky survey, especially at large distances. Fortunately, the SZ
thermal effect, which has a specific spectral signature (positive and negative
intensity peaks at respectively 0.85 and 2 mm with a zero value at 1.38 mm), 
is rather
easy to identify in the framework of a multi-frequency experiment covering
the range 30 to 800\,GHz, with high sensitivities and good angular
resolution such as COBRAS/SAMBA. Using the separation of processes (primordial
CMB, clusters, free-free, ...) which is based on the projection of the
``observed'' signal in different wavebands (Bouchet et al. 1996), the sky
simulation provides us with a recovered $y$ map for the SZ thermal effect
together with a $\Delta T/T$ map.  These recovered maps are used to 
characterize the number of
clusters that could be detected and furthermore they show the ability of the
survey to image the clusters.  \par

In order to find the detected clusters in the recovered $y$ maps, we use the
following algorithm. Above a threshold of {\bf $y=2.\,10^{-6}$}, each maximum 
is
associated with the central peak of the $y$ profile of a resolved cluster.
Assuming spherical symmetry, we compute the radial profile of the
comptonization parameter $y$, by averaging within rings of equal width. We
also compute the integrated $y$ parameter, Y, over the profile for each
recovered cluster. Then, we compare these reconstructed profiles with the
input ones. We thus produce a catalogue of detected clusters giving the
position of the $y$ maxima in the map, together with the ``measured'' central
value $y_0$, and the integrated parameter Y.  The method, which consists of
integrating the signal over rings getting larger and larger, tends to
overestimate the signal if one does not take out a base line which is the
average $y$ value given by undetectable weak clusters.  For rings of constant
$\delta R/R$, both noise and signal decrease as $1/R$ where $R$ is the
distance to the center of the cluster; thus the signal (integrated over rings)
to noise ratio is constant as long as the density follows a $\beta$-profile.
\begin{figure}
\psfig{file=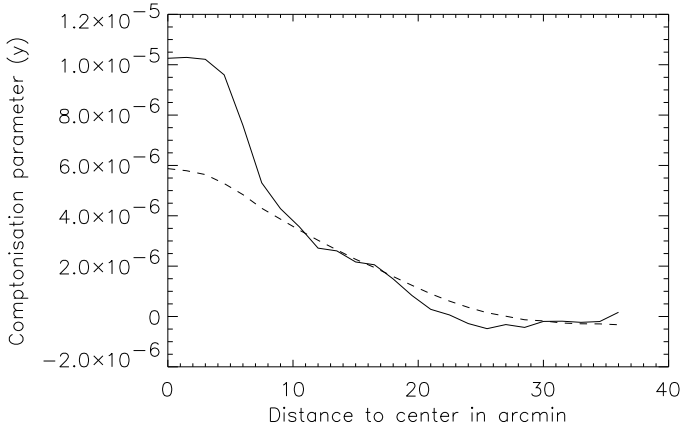,width=\columnwidth}
\caption[ ]{Input (solid line) and recovered (dashed) profiles for a cluster
with high $y$ as a function of the distance to center in arcminutes. The 
integrated $y$ parameter is recovered with better than 2\%} 
\end{figure}
%
%
\begin{figure}
\psfig{file=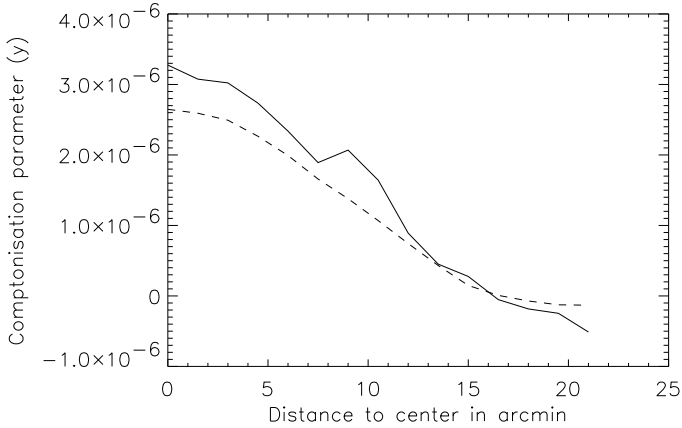,width=\columnwidth}
\caption[ ]{Input (solid line) and recovered (dashed) profiles for a cluster
100 times weaker than typical observed clusters. Here, the intagrated $y$ 
parameter is recovered with an accuracy of 12\%.}
\end{figure}
\cali
\par

Figures 2 and 3 show both input and recovered profiles for respectively a
strong cluster (in terms of the $y$ parameter) and a weak one, very close to
the sensitivity limit of the COBRAS/SAMBA instruments. In both figures we
note that the central part of the cluster suffers from the beam dilution, this
region is therefore not recovered in a satisfactory way by the SZ
observations. On the other hand, the wings of the cluster are well recovered,
they are observable down to $y\approx3.\,10^{-7}$ over about one degree as
indicated in Fig. 2. The comparison between the input and recovered profiles 
shows a good
agreement. In fact, the sensitivity is such that clusters, and specifically
the wings of the profile, can be observed through the SZ thermal effect as far 
as the gas extends; the main limit to such a measurement is likely to be the
confusion limit, due to the overlap of weak clusters in the background.  The 
ability to
detect the wings of the profiles indicates that the SZ effect is a powerful 
tool
to constrain gas accretion models in potential wells of clusters and measure
the virialization radius which gives the spatial extent of a
cluster. Furthermore, more information in the statistical properties of the 
distribution
of the comptonization parameter $y$ can be extracted and used to constrain
evolution models even when confusion sets in.  \par

The X-ray emission and the SZ effect are dominated by different parts of the
cluster; X-rays are mainly associated with the core of the cluster because the
intensity of bremstrahlung emission is proportional to $n_e^2$, whereas the SZ
effect which is proportional to the electron column density is dominated by
contributions coming from the lower density regions where the path length is
longer.  Therefore, the information brought by both X-rays and the SZ effect
are quite different and rather complementary.  \par

\noindent
A comparison between the capabilities of X-ray and SZ measurements for
clusters can be done by using the characteristics of the EPIC camera planned
for the X-ray observatory mission XMM and the COBRAS/SAMBA characteristics.
Due to their good resolution and because they are biaised by the $n_e^2$
dependence, X-ray observations are well adapted to map the core of the
clusters. The EPIC-XMM instruments can map the cluster up to about $7\theta_c$
at a sensitivity of $5\sigma$ reached in 20 hours of integration. This is
illustrated in figure 4
for an A496-like cluster ($T_{e0}=4$ keV at $z=0.2$). The EPIC
instrument will map the clusters (especially the central part) and will give
the temperature distribution, electronic density...  The SZ observations suffer
from a significant beam dilution of the core, because of the lack of
resolution compared with the X-rays measurements, but they resolve the 
wings of
the cluster profile up to the virialization limit (cf. Fig. 2). Figure 4 shows
that the A496-like cluster is mapped up to about $17\theta_c$ with SZ
observation.  Therefore, a combination of X-ray measurements together with SZ
observations will strongly constrain cluster models.
\begin{figure}
\psfig{file=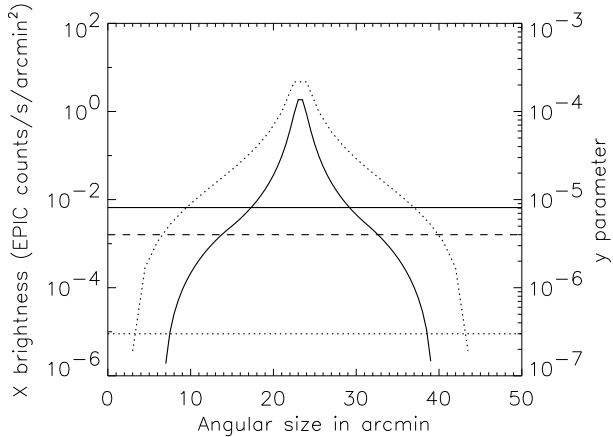,width=\columnwidth}
\caption[ ]{Comparison between X-ray brightness (solid line) and the $y$ 
parameter (dotted line) for a
A496-like cluster ($T_{e0}=4$ keV at $z=0.2$) with XMM-EPIC and COBRAS/SAMBA
resolutions, respectively for the X-rays and $y$ observations.  Left axis:
X-ray brightness in EPIC counts/s/arcmin$^2$; right axis: $y$ paramter, versus
angular size in arcmin. The solid horizontal line stands for the X-ray
background, the sensitivity limit of the XMM instrument for an integration
time of 20 hours is indicated by the dashed horizontal line whereas the dotted
line is for the COBRAS/SAMBA sensitivity limit.}
\end{figure}
\cali
\par

A high sensitivity SZ sky survey like the one that can be carried out by a
mission like COBRAS/SAMBA will give the best catalogue of distant clusters and
will be used to define follow-up observations with X-ray observations which
require a long integration time to study a distant cluster.  \par

One can characterize the clusters in terms of their measurable parameters:
angular core radius $\theta_c$, central value of the comptonization parameter
$y_0$ or the integrated parameter Y, which combines the two previous
quantities.  Assuming a King profile, this parameter is directly related to
the mass of the gas, the redshift $z$ and the temperature of the clusters, by
the following expression (for a flat universe): 
\[
\int\frac{y~d\Omega}{10^{-4}~\mbox{arcmin}^2}=0.43h^2 \left(\frac{M_G}
{10^{14}M_{\odot}}\right) \left(\frac{kT_e}{10~keV}\right)
\]
\begin{equation}
\frac{(1+z)^3}{(\sqrt{1+z}-1)^2} .
\label{eq:yint}
\end{equation}
In terms of observable parameters, Y is given by ${\mbox
Y}=64\,y_0(\theta_c /\mbox {arcmin})^ 2$. We expect that the extraction of the
clusters will depend mainly on the value of Y, because the sensitivity of the
SZ profile does not decrease with radius as long as the gas density decreases
as $R^{-2}$.{\it  Thus, the sensitivity required to detect a cluster is 
controled by the integrated emission and not by the peak brightness.}  \par
We use the extraction method described above, applied to the simulated input
and recovered maps and derive the completeness of the obtained catalogue 
of
resolved clusters as a function of their Y.
\begin{figure}
\psfig{file=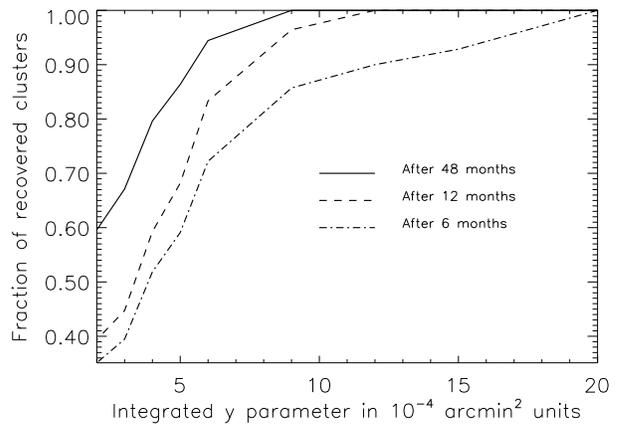,width=\columnwidth}
\caption[ ]{Fraction of the recovered clusters from a  recovered map after
component separation, for three sensitivities as a function of the integrated
Y comptonization parameter. The nominal mission of COBRAS/SAMBA is 14 months
of observations.}
\end{figure}
\cali
A systematic extraction of the clusters is done for three noise configurations
(the nominal observing time for COBRAS/SAMBA mission is about one year). Both
48 and 12 months of observations enable the detection of at least 95\% of the
clusters with Y $>9.\, 10^{-4}\,\mbox {arcmin}^2$. For the 12 months 
configuration, 70\% of the clusters with Y
$>5.\,10^{-4}\,\mbox {arcmin}^2$ and $y_0>3.\,10^{-6}$ are
recovered. For such clusters, the integrated $y$ parameter is recovered, on 
average, with an accuracy of about 30\% (clusters shown in figures 2 and 3, the
accuracies are respectively 1.4\% and 12\%). Decreasing the noise level leads 
obviously to a deeper survey as indicated in the plot (Fig. 5).  If we 
increase the noise level to a noise
corresponding to a 6 months mission, even some of the strongest clusters are
not detected. In fact, we recover only 85\% of clusters for which ${\mbox Y}
>10^{-3}\,\mbox {arcmin}^2$, and the extraction gets worse for clusters with
${\mbox Y}=5.\,10^{-4}\,\mbox {arcmin}^2$.  
\par
By using the De Luca et al. (1995) source counts model, we expect that a
mission like COBRAS/SAMBA could detect about $7.\,10^{3}$ resolved clusters
with ${\mbox Y}>9.\,10^{-4}\,\mbox {arcmin} ^2$ and a completeness better than
95\%, and about $10^{4}$ clusters with ${\mbox Y}>5.\, 10^{-4}\,\mbox
{arcmin}^2$ and completeness better than 68\%. These numbers are very model
dependent and could increase by a factor 3 for cosmological models with low
$\Omega_0$ due to the existence of  distant clusters ($z>0.5$) and their 
contribution to the counts as in the model of Barbosa et al. 1996.
\section{Peculiar velocity measurement}
Given the radial peculiar velocities at known positions and  
under the assumption of a potential velocity flow around our galaxy, one can 
extract the
velocity field and derive the bulk velocity $V_R$ which is the average of the
local velocity field smoothed over a window function of scale $Rh^{-1}$
Mpc. One can get the peculiar radial velocities using the redshift surveys and
some relationships, giving the distances to the objects, such as the
Faber-Jackson (1976) or Tully-Fisher (1977) relations.  Several authors have
measured the bulk velocities averaged over different volumes (a good
approximation at very large scales), Dressler et al. (1987) found $V_{60}=599
\pm104$ km/s, Courteau et al. (1993) measured $V_{40}=335 \pm38$ km/s and
$V_{60}=360\pm40$ km/s, Willick et al. (1996) computed the bulk velocity 
using POTENT they found $V_{60}=20 - 300$ km/s.  The
main difficulty from which this method suffers is that it requires very
reliable distance indicators. Therefore, the relative error in determining the
peculiar velocity increases with distance.

The determination of the peculiar velocity from SZ measurement is promising,
since its accuracy is distance independent. If indeed one can detect the 
kinetic SZ effect for
a number of clusters as large as $10^{4}$ (as possible with the
COBRAS/SAMBA mission), this will provide unique statistical information on the
velocities.  \par
\subsection{ Calculation method and geometrical filter}
The radial peculiar velocity of the cluster is given by combining the
measurements of both kinetic and thermal SZ effects (Sect. 2.2, Eq. 3).  
The components separation takes advantage of the spectral signatures of
the different astrophysical contributions (free-free, synchroton, clusters of
galaxies, ...), which are taken into account in the sky simulation (Bouchet et 
al. 1995), and of the characteristics of the instrument to give
recovered maps of the astrophysical processes. In particular, the
$\Delta T/T$ map, obtained after the components separation, includes both
primordial CMB fluctuations and $(\Delta T/T)_{SZ}$ because they have the 
same
spectral signature.  The $(\Delta T/T)_{SZ}$ is thus contaminated by the
primordial temperature fluctuations of the CMB, this contamination beeing
responsible for an error in the determination of the peculiar radial velocity
of a cluster when one uses a method based on the combination of measurements
of both SZ effects. The aim is to have the smallest error on the velocity,
this requires a good measurement of $(\Delta T/T)_{SZ}$.  \par The induced
temperature fluctuations due to the SZ kinetic effect have angular sizes
smaller than one degree, whereas the CMB spectrum exhibits a ``Doppler'' peak 
at
about one degree. Within this framework, a good measurement of the SZ kinetic
effect (and thus the peculiar velocity) must be a compromize between on the
one hand maximizing the signal by integrating over
a large beam, and on the other hand minimizing the spurious contribution from
the CMB.  \par

The peculiar radial velocity of a cluster is given by
$$v_r=\frac{ck}{m_e c^2}\times T_e\frac{\Delta T/T}{y}.$$ The relative error
in the velocity is thus expressed as follows: 
\bequ \frac{\delta v_r}{v_r} =
\frac{\delta A}{A} + \frac{\delta T_e}{T_e}, 
\fequ 
where $A=\frac{\Delta T/T}{y}$.  The error due to the CMB contamination 
appears in the
measurement of the temperature fluctuation $\Delta T/T$ and it is measured
using a spatial filter that optimizes the signal to ``noise'' ratio (the main
``noise'' being the primary CMB). The $\Delta T_e/T_e$ term is the error due
to the uncertainty on the intracluster gas temperature determination. It
should be derived from X-ray data. Hereafter, we do not include 
it in the evaluation of the $\frac{\delta v_r}{v_r}$.  
\par 
To minimize the
contamination of the CMB, we construct a spatial filter over which we
calculate a variation of both $\Delta T/T$ and $y$ on a single cluster. The
spatial filter used hereafter computes the difference between the mean values
of $\Delta T /T$ and $y$ in the central part of the cluster and their mean
values taken in a ring around the peak. This filter has three free parameters,
one is the radius of the central disc and the two others the inner and outer
ring radii.  \par

If the noise due to the spurious signals is dominated by a single component of
a known power spectrum, the spatial filter can be optimized (Haenhelt \&
Tegmark 1996). In practice, several sources contribute to the noise including
some non gaussian ones like the confusion of weaker clusters. Furthermore, we
adress the problem of the optimization of the filter using only the 
informations included in the data. The shape
and size of the optimum filter depend on the extent of the cluster and its
density profile.  We thus empirically optimize the simple filter in the case
of one cluster model (King profile) and one primordial CMB spectrum (standard 
CDM with
$\Omega_b=0.05, h=0.5$ and $ \Omega_0=1$) to evaluate the overall accuracy of
the velocity determination.  The errors are dominated by the modes of the CMB
power spectrum of wavelengths comparable or greater than the core radius.
\par

The optimum parameters of the filter are obtained by minimizing $\delta v_
{rms}$,where the {\it rms} velocity is obtained after many realizations of the
millimeter and sub-millimeter sky including the astrophysical 
contributions
due to CMB, foregrounds and clusters of galaxies and the instrumental 
noise.
\par
\subsection{Results}
Our goal is to have a geometrical filter that could be applied for a wide
range of cluster sizes and derived directly from observations. We find that 
the optimized spatial filter has a
central disc, associated with the peak of the cluster, corresponding to the
region where $y$ is greater than 70\% of its maximum value $y_0$. For the
ring, the best compromize is obtained for an inner radius $R_{in}=0.5\times
FWHM$ and a width $\Delta R=2$ pixels.  Hereafter, we use the same optimized
filter parameters for all cluster sizes. We check that varying both $\Delta R$,
in a
range of 1 to 3 pixels, and $R_{in}$, in range of 0.5 to $1.5\times FWHM$,
introduces an error of a few percent only in the velocity determination.  \par
\begin{figure}
\psfig{file=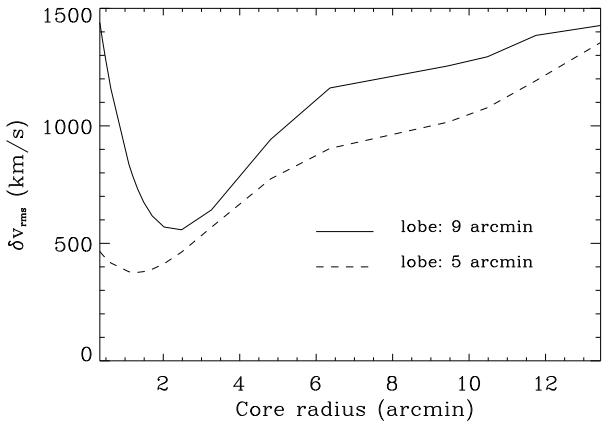,width=\columnwidth}
\caption[ ]{$\delta v_{rms}$ due to the contamination by CMB fluctuations as a
function of the core radius of the cluster, for a cluster with $y\approx
10^{-4}$. Solid line: convolution with a 9 arcminutes beam; dashed line: for a
5 arcminutes beam.}
\end{figure}
\cali
Figure 6 displays the $\delta v_{rms}$ obtained with the optimized filter (for
a cluster with $y_0\simeq 10^{-4}$) as a function of the core radius of the
cluster. It shows that the {\it rms} velocity decreases for decreasing sizes
of the clusters.  We expect that the modes of the CMB power spectrum which
contribute mostly to the measurement of the radial velocity of the clusters
are those with wavelengths comparable to the size of the ring of the
geometrical filter, which is given by $2 \times (R_{in} + \Delta R) = (3.5
\times \theta_c + 6')$. Thus, the contamination of the CMB increases with the
size, as more contribution from the first Doppler is included. This
explains the rise of $\delta v_{rms}$ from $2-3$ arcminutes upwards. For
clusters with core radii smaller than 2 arcminutes, the beam dilution leads to
a fast degradation of the velocity determination accuracy.  We use gaussian
beams of 9 and 5 arcminutes in order to evaluate the effect of the
beam dilution. As expected, Figure 6 shows that the accuracy improves for a
smaller beam, for all the cluster sizes but the effect is striking for core 
radii smaller than 1.5 arcminute.  \par

We check that the velocity determination accuracy for the largest clusters
does not vary when we increase the noise since, in this range, the CMB
contamination is dominant. For small clusters, we also find that the 
$\delta
v_{rms}$ does not increase by more than 30\% when we increase the noise by a
factor 3.  \par

\noindent
We compare the accuracy determination for a rich cluster ($y_0\simeq10^
{-4}$) and a weak one ($y_0\simeq 3.\,10^{-5}$). Obviously, the evaluation of
the velocity is more accurate for the richest clusters. In fact, we find that,
as expected, the velocity uncertainty scales as $1/y_0$. This enables to
derive the accuracy in the velocity determination for each cluster, knowing
its core radius $\theta_c$ and its central $y$ parameter, $y_0$. In the 
framework
of a particular cluster model (here Bartlett \& Silk 1994), one can deduce
$\delta v_{rms}$ as a function of the cluster mass. 
 \par
Haenhelt \& Tegmark (1996) have discussed in detail the optimized spatial
filter for CMB contamination only, for various cosmological parameters. Our
simulations, which take into account a more realistic noise together with the
limitations due to other astrophysical contributions (dust, free-free, ...),
confirm that the accuracy they found can be achieved when the other
sources of noise are taken into account.
\par
An additional uncertainty, noted $\delta T_e/T_e$ in Eq. 5, comes from 
the determination of the temperature of the intracluster medium from X-ray
observations. This uncertainty is strongly dependent on the observed cluster
and the characteristics of the instruments. With an instrument such as 
XMM-EPIC, 
the temperature of nearby clusters will be determined with a very high 
accuracy ($\simeq$ 5\%) for more distant clusters the uncertainty could be as
high as 10\%. The results of the uncertainty on peculiar velocity evaluation,
given Fig. 6, have been obtained neglecting the contribution of the uncertainty
on the temperature
$T_e$. This contribution, which is not dominant, should be added according to 
the accuracy of the specific X-ray data used.
\section{Velocity dispersion measurements}
Our simulations show that the measurement of the peculiar cluster velocity is
marginally possible only for the strongest clusters, in terms of their $y$
parameter ($y_0> 10^{-4}$). Furthermore, the number of such clusters over the
sky is too small to give very useful statistical information.  Meaningful
measurements can only come from a statistical analysis, one beeing the {\it 
rms}
velocity dispersion, the other beeing the bulk velocity.  \par

The measurement of the $rms$ radial velocity $(v_{rms})_{measured}$ of a
cluster is the combination of its real velocity dispersion $v_{rms}$ and of
the velocity uncertainty due to the background contribution $(v_{rms})_{
instr}$. We therefore have:
$$
(v_{rms})_{measured}^2 = v_{rms}^2 + (v_{rms})_{instr}^2
$$
The error in the determination of the $rms$ velocity on N clusters decreases
like $1/ \sqrt N$.  Therefore, it is necessary to measure the SZ effects on a
large number of clusters in order to have the smallest error in
$(v_{rms})_{measured}$.  Nevertheless, this kind of information is only
partially relevant since the main difficulty is that such a measurement
requires a good evaluation of $(v_{rms})_{instr}$ which is not easily
achieved.  \par

Over large scales, another accessible piece of statistical information from
the peculiar velocity of the clusters is the bulk velocity averaged over given
volumes.  In the specific model of this paper: De Luca et al. (1995) for the
cluster counts and Bartlett \& Silk (1994) for the evolution, and taking into
account the limits due to the sensitivity of the experiment, one can derive
the number of observed clusters per unit solid angle and intervals of redshift
and mass. For a given mass and redshift the observable parameters
($\theta_c$, $y_0$) are kown and using the results of section 5.2. (Fig. 6), 
one can
evaluate the accuracy $\delta v_{rms}$ of the radial peculiar velocity for
each class of clusters.  
\par\noindent
In a given volume containing N clusters with individual peculiar velocities
$v_i$ and accuracies $\sigma_i$, the best estimate of the bulk velocity $V$ is
given by the mean weighted velocity:
$$
V=<v>=\frac{\displaystyle\sum_{i=0}^N\frac{1}{\sigma_i^2}v_i}{\displaystyle
\sum_{i=0}^N\frac{1}{\sigma_i^2}}
$$
One can compute the overall accuracy $\sigma$ in the same volume, which is
given by:
$$
\sigma^2=\frac{\displaystyle\sum_{i=0}^N\frac{1}{\sigma_i^2}(v_i-V)^2}{(N-1)
\displaystyle\sum_{i=0}^N\frac{1}{\sigma_i^2}}
$$
We define a local volume $\mbox{V}_{loc}$ by the volume being within the 
redshift range $0<z<0.05$ and which corresponds to a radius of $150h^{-1}$ Mpc.
We also define, up to redshift $z=0.9$, volumes equal to $\mbox{V}_{loc}$
($\mbox{V}_{loc}\simeq10^7\,h^{-3}\mbox{Mpc}^{3}$).
We compute the number of clusters in these volumes by integrating the counts
over masses and redshifts. For each cluster, we also compute the observables 
$\theta_c$ and $y_0$. Using the results given Fig. 6, to which we add a 
conservative uncertainty of 20\% due to the estimate of the intracluster
temperature, and together with the computed values of $\theta_c$ and $y_0$, we
derive the individual accuracy in the pecliar velocity determination for
each cluster. We thus compute the overall accuracy $\sigma$ in each volume.
\par
Our estimates are summarized in Table 1. We find that the overall
accuracy in the local volume is about 60 km/s. It goes through a minimum around
$z=0.1$ but it lower than 100 km/s for $0.5<
z<0.7$. At higher redshifts, the overall accuracy in the peculiar velocity
determination reaches about 200 km/s.
\begin{table}[h]
\caption{Overall accuracy $\sigma$ in the peculiar velocity measurement for 
different redshift ranges.}
\begin{center}
\begin{tabular}{|l|ll|}
\hline \hline $z$ & $\Omega$ (sr) & $\sigma$ (km/s) \\ \hline \hline 
0-0.05 & 4$\pi$ & 60 \\
0.05-0.1 & 2.2 & 12 \\
0.1-0.3 & 0.157 & 28 \\
0.3-0.5 & 0.098 & 34 \\
0.5-0.7 & 0.096 & 94 \\
0.7-0.9 & 0.11 & 223 \\
\hline
\end{tabular}
\end{center}
\end{table}
\par
The measurements of the the bulk velocities from the peculiar
velocities  of individual clusters obtained using the combination of the 
SZ thermal and kinetic effects can
be done with an overall accuracy better than 100 km/s up to $z=0.7$. We have 
thus shown that a survey of the sky in the SZ effects followed by a redshift
survey of a large number of detected clusters gives the possibility of mapping 
the velocity fields in the Universe on very large scales ($>100
h^{-1}$ Mpc) with a good accuracy. In the assumption of a potential
flow and using some reconstruction method such as POTENT (Bertschinger \& 
Dekel 1989), one can then derive the full three-dimensionnal
velocity field and thus the density field which in turn allows comparisons
with the one traced by galaxies. Therefore, the 
application of the SZ effect
measurements to the evaluation of the peculiar velocity of clusters of 
galaxies
should become a very useful tool to test and constrain theories of structure 
formation and evolution.
\section{Conclusions}
Future high sensitivity and high resolution CMB experiments will have to
remove the contributions from various foregrounds, including 
that generated
through the Sunyaev-Zeldovich effect. Fortunately this effect can be easily
separated through its spectral signature. Conversely, this offers the exciting
prospect of creating a large catalogue of clusters selected entirely via this
effect, thereby allowing detections at high redshifts. In addition,  using
both thermal and kinetic SZ effects, one should
then be able to map the very large scale velocity field.

In order to make rather realistic predictions on the potential of such
experiments, we have used complete sky simulations developped to assess the
capabilities of a multi-frequency (30 to 800\,GHz) space mission dedicated
to the observation of the CMB between a few arcminutes to ten degrees, with a
sensitivity close to $\Delta T/T\simeq2.\,10^{-6}$ (Bouchet et al. 1995). 
These simulations take
into account both expected astrophysical components and the instrumental
characteristics of the future space mission COBRAS/SAMBA.

We have shown that a COBRAS/SAMBA like experiment will indeed give a fairly
complete catalogue of more than $10^4$ resolved clusters, up to a redshift of
one or more. In the case of resolved clusters with central comptonisation
parameter $y_0 >2.\,10^{-6}$, it is possible to reconstruct their $y$
profiles up to one degree radius for the strongest ones, showing that the main
limitation in the observation of the outer parts of the cluster profile will be
due to the confusion with weaker clusters.  
\par
Measuring the peculiar velocities of clusters is possible (Eq. 3) when we
combine both SZ thermal and kinetic effects. We have constructed and optimized
a geometrical filter for this purpose. Our results, taking into account all
the astrophysical and instrumental contaminants in addition to the CMB
emission, confirm the Heanhelt \& Tegmark (1996) results for which CMB was the
only spurious signal. We have also shown that measuring the peculiar
velocities, for a large number of clusters up to
$z=0.5$ gives overall accuracies on bulk velocities, computed in regions of
typical dimension $100\,h^{-1}$ Mpc, better than 100 km/s. This suggests that 
it will therefore be 
possible to map the density field of the Universe using SZ velocity 
measurement.  
\par 
The overall accuracy
at higher $z$ depends more strongly on the cluster counts which are not
constrained strongly. In any case, the specific model we used here, which 
takes into account the evolution of the number of clusters gives, if anything,
an underestimate of the counts compared with no-evolution models.
\begin{acknowledgements}
The authors thank F.X. D\'esert and M. Lachi\`eze-Rey for useful discussions,
J.R Bond for providing us with one of the SZ maps we analyzed  and an 
anonymous referee for useful comments.
\end{acknowledgements}
\section{APPENDIX A: $y$ profile for a resolved cluster}
In the Rayleigh-Jeans part of the spectrum, the SZ thermal effect is given by:
\[
\Delta T_{RJ}=-2T_{CMB}\frac{k\sigma_T}{m_ec^2}\int n_eT_edl.
\]
Using the assumptions of isothermality and spherical symmetry to describe the
gas distribution, we have:
\[
\Delta T_{RJ}=-2T_{CMB}\frac{k\sigma_T}{m_ec^2}n_{e0}T_{e0}D_A\int f_nd\zeta,
\]
with:
\[
f_n=\left(1+\frac{\theta^2+\zeta^2}{\theta^2_c}\right)^{-\frac{3\beta}{2}}.
\]
We have $R_{max}=20~R_c$ which we take as an integration limit for the
integral above.  \par

 To compute the integral up to $R_{max}$, we split the integration to infinity
into two parts and write:
$$I_{0,R}=I_{0,\infty}-I_{R,\infty}$$,
with:
\[
I_{0,R}=\frac{1}{D_A}\int^{R_{max}}_0 \left(1+\frac{\theta^2}{\theta_c^2}+
\frac{l^2}{r^2_c}\right)^{-\alpha}dl,
\]
\[
I_{0,\infty}=\frac{1}{D_A}\int^{\infty}_0 \left(1+\frac{\theta^2}{\theta_c^2}+
\frac{l^2}{r^2_c}\right)^{-\alpha}dl,
\] 
\[
I_{R,\infty}=\frac{1}{D_A}\int^{\infty}_{R_{max}}
\left(1+\frac{\theta^2}{\theta_c^2}+ \frac{l^2}{r^2_c}\right)^{-\alpha}dl.
\]
We finally find:
\[
I_{0,R}=\frac{1}{2}\theta_c\sqrt{\pi}\frac{\Gamma (\alpha- 1/2)}{\Gamma
(\alpha)}\left(1+\frac{\theta^2}{\theta^2_c}\right)^{1/2-\alpha}+
\theta_c\frac{\omega^{1-2\alpha}_{max}}{1-2\alpha}\]

\[-\frac{\theta_c}{2}\left(1+
\frac{\theta^2}{\theta^2_c}\right)\frac{\omega^{-(1+2\alpha)}_{max}}{1+2\alpha
},
\]
where:
$$
\omega^2_{max}=1+\frac{R^2_{max}}{R^2_c}.
$$
Therefore, the temperature variation $\Delta T_{RJ}$ is written as:
\[
\Delta T_{RJ}=-2T_{CMB}\frac{kT_{e0}}{m_ec^2}\sigma_T n_{e0} D_A I_{0,R}.
\]
We have seen in, Sect. 2.1, the relation between the $y$ parameter and
temperature variation which gives:
\[
y=-\frac{1}{2}\frac{1}{f(x)}\frac{\Delta T_{RJ}}{T_{CMB}}.
\] 
The profile of the $y$ parameter is thus derived directly from the expression
of the temperature variation, and the $\Delta T/T$ profile is obtained using
Eq. 2, Sect. 2.2.
\end{document}